\documentclass[10pt, conference, compsocconf]{IEEEtran}
\usepackage{amssymb}
\usepackage[usenames,dvipsnames,svgnames,table]{xcolor}		%-To write text in color
\usepackage{url}		%-Web addresses
\usepackage{epstopdf}

\usepackage{fixme}
\fxsetup{
  status=draft, % change to final to hide all remaining fixmes
  multiuser,
  theme=color,
  inlineface=\bfseries,
  envface=\bfseries
}
\FXRegisterAuthor{ne}{ane}{NE} % Norbert Eicker

\usepackage{cite}
\usepackage{graphicx}
\ifCLASSINFOpdf
\else
\fi

\usepackage{listings}

\definecolor{mygreen}{rgb}{0,0.6,0}
\definecolor{mygray}{rgb}{0.5,0.5,0.5}
\definecolor{mymauve}{rgb}{0.58,0,0.82}

\usepackage{caption}
\usepackage[font=footnotesize]{subfig}
\usepackage{url}
\begin{document}
%
% paper title
% can use linebreaks \\ within to get better formatting as desired
\title{Application performance on a 
Cluster-Booster system}

% author names and affiliations
% use a multiple column layout for up to two different
% affiliations

\author{\IEEEauthorblockN{Anke Kreuzer, Norbert Eicker}
\IEEEauthorblockA{J\"{u}lich Supercomputing Centre (JSC) \\ 
		Institute for Advanced Simulation (IAS)\\
        Forschungszentrum J\"{u}lich GmbH \\  
		52425 J\"{u}lich, Germany}
\and
\IEEEauthorblockN{Jorge Amaya}
\IEEEauthorblockA{Katholieke Universiteit Leuven \\ 
	Centre for Mathematical \\ 
	Plasma-Astrophysics \\
	Mathematics Department \\
	3001, Leuven, Belgium}
\and
\IEEEauthorblockN{Estela Suarez*}
\IEEEauthorblockA{J\"{u}lich Supercomputing Centre (JSC) \\ 
		Institute for Advanced Simulation (IAS)\\
        Forschungszentrum J\"{u}lich GmbH \\  
		52425 J\"{u}lich, Germany \\
	Email*: e.suarez@fz-juelich.de}	%Corresponding Author
}

\maketitle

\begin{abstract}
The \textit{DEEP} projects have developed a variety of 
hardware and software technologies aiming at improving 
the efficiency and usability of next generation
high-performance computers. They evolve around an innovative
concept for heterogeneous systems: the Cluster-Booster 
architecture. In it, a
general purpose cluster is tightly coupled to a many-core system
(the Booster). This modular way of integrating heterogeneous 
components enables applications to freely choose the kind 
of computing resources on which it runs most efficiently. Codes might
even be partitioned to map specific requirements of
code-parts onto the best suited hardware. 
This paper presents
for the first time measurements done by a real world scientific
application demonstrating the performance gain achieved with
this kind of code-partition approach. 
\end{abstract}

\begin{IEEEkeywords}
Exascale;  Architecture; Cluster-Booster architecture; Modular Supercomputing; Co-design

\end{IEEEkeywords}

% For peer review papers, you can put extra information on the cover
% page as needed:
% \ifCLASSOPTIONpeerreview
% \begin{center} \bfseries EDICS Category: 3-BBND \end{center}
% \fi
%
% For peerreview papers, this IEEEtran command inserts a page break and
% creates the second title. It will be ignored for other modes.
\IEEEpeerreviewmaketitle

\section{Introduction}
\label{sec:intro}
The high-performance community is addressing multiple 
challenges to provide industrial and scientific 
users with suitable and efficient Exascale 
systems. Huge power consumptions, much faster growth of computing 
capabilities than memory and I/O bandwidth (the so-called memory wall), 
and higher hardware failure rates expected in 
such huge systems, are some examples. 
Also extreme concurrency and the integration of heterogeneous
computing resources do affect the programmability of a 
system, since both require specific
code adaptations to fully exploit the capabilities 
of the platform. %All these aspects have a direct
%impact on the scientific throughput that can be extracted 
%from Exascale systems.

The \textit{DEEP projects}~\cite{DEEPweb} are a series 
of three EC-funded projects (DEEP, \mbox{DEEP-ER}, 
and \mbox{DEEP-EST}) performing research addressing the 
Exascale computing challenges. The first member of this family 
(\textit{DEEP: Dynamical Exascale Entry Platform}) introduced 
a new heterogeneous supercomputer architecture: the 
Cluster-Booster concept~\cite{eicker:PARS}, aiming at 
increasing the scalability and energy efficiency of 
cluster systems, while
keeping their programmability and flexibility. 
DEEP built a first hardware prototype, including 
a complete software stack with resource management, 
scheduler, programming environment, and performance analysis 
tools~\cite{eicker:CCPE}. \textit{\mbox{DEEP-ER} (DEEP - Extended Reach)} 
extended the Cluster-Booster architecture implementing a multi-level 
memory hierarchy, acting as a basis for a complete I/O 
and resiliency software stack. Finally, the recently
started project \textit{\mbox{DEEP-EST} (DEEP -- Extreme Scale 
Technologies)} generalises the Cluster-Booster concept 
introducing the so-called \textit{Modular Supercomputing architecture}~\cite{Suarez:844072}.

All three projects follow a stringent \textit{co-design} 
strategy, using full-fledged scientific applications to
guide and strongly influence the design and implementation of system 
hardware and software. The applications requirements, identified 
by detailed analysis, guided   
all the project's developments. The selected codes 
have also been adapted to the Cluster-Booster platform and 
served as a measure to validate and benchmark the hardware and
software technologies implemented.

This paper describes the Cluster-Booster architecture,
its second-generation prototype (\mbox{DEEP-ER} prototype), 
the software environment, and the advantages that the
concept brings to applications exemplified by some of 
the results achieved within the \mbox{DEEP-ER} project.
Section \ref{sec:arch} 
presents the \mbox{DEEP-ER} system architecture, 
including the underlying Cluster-Booster concept, 
%(\ref{sec:CBconcept})
the specific hardware configuration of the 
\mbox{DEEP-ER} prototype, 
%(\ref{sec:sysconf})
and its memory hierarchy and technologies. 
%(\ref{label{sec:mem}}, \ref{sec:nam}} 
The software stack is explained in section~\ref{sec:sw}, 
including the programming environment 
already introduced in the predecessor DEEP project, 
%{\ref{sec:progenv}, \ref{sec:ompss})
and a summary of the \mbox{DEEP-ER} 
I/O and resiliency software developments.
%(\ref{sec:io}, \ref{sec:resil})  
The application used to evaluate the Cluster-Booster
architecture is shortly described 
in section~\ref{sec:results}, together with the
results achieved distributing it over both 
parts of the \mbox{DEEP-ER} prototype. 
Finally, the conclusions of the 
paper are summarised in section~\ref{sec:concl}.

\section{System Architecture}
\label{sec:arch}

Cluster computing enables to build high-performance systems 
benefiting from the lower cost of commodity of the 
shelf (COTS) components. 
Traditional, homogeneous clusters are built by connecting a 
number of general purpose processors (e.g. Intel Xeon, 
AMD Opteron, etc.) using a high speed network. The limitation of 
this approach lies on the relatively high power consumption 
and cost per performance of general purpose processors, 
which makes a large scale homogeneous system made of this kind 
of processors extremely power hungry and costly. 

The overall energy and cost efficiency of a cluster 
can be improved by 
adding accelerator devices (e.g. many-core processors or 
general purpose graphic cards, GPGPUs), which provide 
higher Flop/s performance per Watt. 
Standard heterogeneous 
clusters are built by attaching one or more accelerators 
to each node. However, this \textit{accelerated 
node} approach presents some caveats. An important one is 
the combined effect of the accelerators' dependency on the 
host CPU and the static arrangement of hardware resources, 
which limit the accessibility to the accelerators for other 
applications than the one running on the host CPU. Furthermore, 
both CPU and accelerator have to compete for the scarce 
network bandwidth in this concept.

\subsection{Cluster-Booster concept}
\label{sec:CBconcept}
\begin{figure}[!t]
	\centering
	\includegraphics[width=\columnwidth]{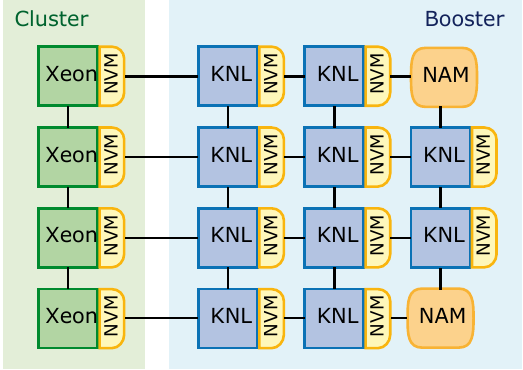}
	\caption{\label{fig:CBarch_scheme} 
	Sketch of the Cluster-Booster architecture 
	in its second generation 
	implemented in the \mbox{DEEP-ER} project 
	(KNL: Knights Landing; NVM: non-volatile memory; 
	NAM: network attached memory)}	
\end{figure}

The \textit{Cluster-Booster architecture} (sketched in  
figure~\ref{fig:CBarch_scheme}) integrates heterogeneous computing 
resources at the system level. Instead of plugging accelerators 
into the node attaching them directly to the CPUs, 
they are moved into a stand-alone
cluster of accelerators that has been named \textit{Booster}. 
These accelerators can act autonomously and communicate 
directly with each other through a high-speed network, not 
needing any host node. Leveraging this feature, full codes 
with intensive internal communication can run on the 
Booster alone, without employing any Cluster node in their execution.

%\begin{figure}[!t]
%\centerline{
%\subfloat[Sketch of the Cluster-Booster
%  architecture as implemented in the \mbox{DEEP-ER} project 
%  (KNL: Knights Landing; NVM: non-volatile memory;
%  NAM: network attached memory]
%  {\includegraphics[height=4cm]{pdf/CBarch_scheme.pdf}
%\label{fig:CBarch_scheme}}
%\hfil
%\subfloat[Picture of the \mbox{DEEP-ER} prototype, at JSC]
%  {\includegraphics[height=4cm]{pdf/deeper_proto.pdf}
%\label{fig:deeper_proto}}}
%\caption{Cluster-Booster architecture in \mbox{DEEP-ER}}
%\label{fig:CBarch}
%\end{figure}

The Booster is attached to a standard HPC Cluster 
via a high-speed network. This connection, together 
with a uniform software stack running over both parts 
of the machine (see section~\ref{sec:sw}), enables 
Cluster and Booster acting together as a unified system. 
This opens up new prospects for application developers,
who have full freedom to decide how they distribute their 
codes over the system. For example, code sections 
requiring high-single thread performance and/or 
large memory capacity will run best on the Cluster 
side, while well-parallelized and vectorised code parts 
will profit from the highly-scalable, energy efficient 
Booster. This is demonstrated in section~\ref{sec:results},
which presents measurements of an application that 
benefits from the Cluster-Booster approach.

In contrast to accelerated clusters, the Cluster-Booster 
concept poses
no constraints on the
combination of CPU and accelerator nodes that an 
application may select, since resources are reserved 
and allocated independently. This has two important effects: 
Firstly, each application can run on an optimal 
combination of resources and achieve maximum performance. 
Secondly, all resources can be put to good use by a
system-wide resource manager. The latter allows 
combining the set of applications in a complementary way, 
increasing throughput and efficiency of use for the overall 
system. In the course of the DEEP project, major efforts 
were put into the extension of batch-system 
capabilities~\cite{prabhakaran}.

In DEEP-ER, additional memory components have been 
added to the Cluster-Booster system, 
including the NAM and the NVMe presented 
in section~\ref{sec:sysconf}.

\subsection{Prototype hardware configuration}
\label{sec:sysconf}

\begin{figure}[!t]
	\centering
	\includegraphics[width=0.5\columnwidth]{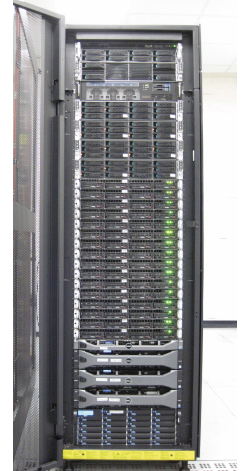}
	\caption{\label{fig:deeper_proto} 
	Picture of the \mbox{DEEP-ER} prototype, at JSC}
\end{figure}

The first prototype of the Cluster-Booster concept 
was designed and built in the course of the \textit{DEEP} 
project~\cite{DEEPweb}. The DEEP prototype consisted of 
128~Cluster nodes (Intel Xeon, Sandy Bridge generation), 
and 384~Booster nodes (Intel Xeon Phi, Knights Corner - KNC generation). 
The different network technologies on Cluster (InfiniBand) and
Booster (EXTOLL) made it necessary to use bridge-nodes
between the two parts of the system. These where responsible
both for transferring messages and for remote-booting the KNC 
nodes from the network, since these were not designed as stand-alone processors.
%While all system software layers have proven full functionality
%on software development vehicles, operational issues of the 
%prototype hardware prevented running scaling 
%measurements of applications on the DEEP system.

The successor --~called the \textit{\mbox{DEEP-ER} prototype}~-- 
(figure~\ref{fig:deeper_proto}) is the second 
generation of the same architecture and was installed 
at JSC in 2016. It consists of 16~Cluster nodes and 8~Booster 
nodes; the configuration is detailed in 
table~\ref{tab:hwconfig}. 
Given the size of the system and the strong focus of the 
\mbox{DEEP-ER} project on software development, the prototype 
construction was kept as simple as possible 
employing off-the-shelf, air-cooled hardware components. 
Cluster and Booster modules are integrated in a single,
standard 19'' rack, which also holds the storage system (one
meta-data, two storage servers and 57~TB of storage on 
spinning disks).

\begin{figure}[!htbp]
  \centering
  %\subfloat[Bandwidth]{
  \includegraphics[width=\columnwidth]{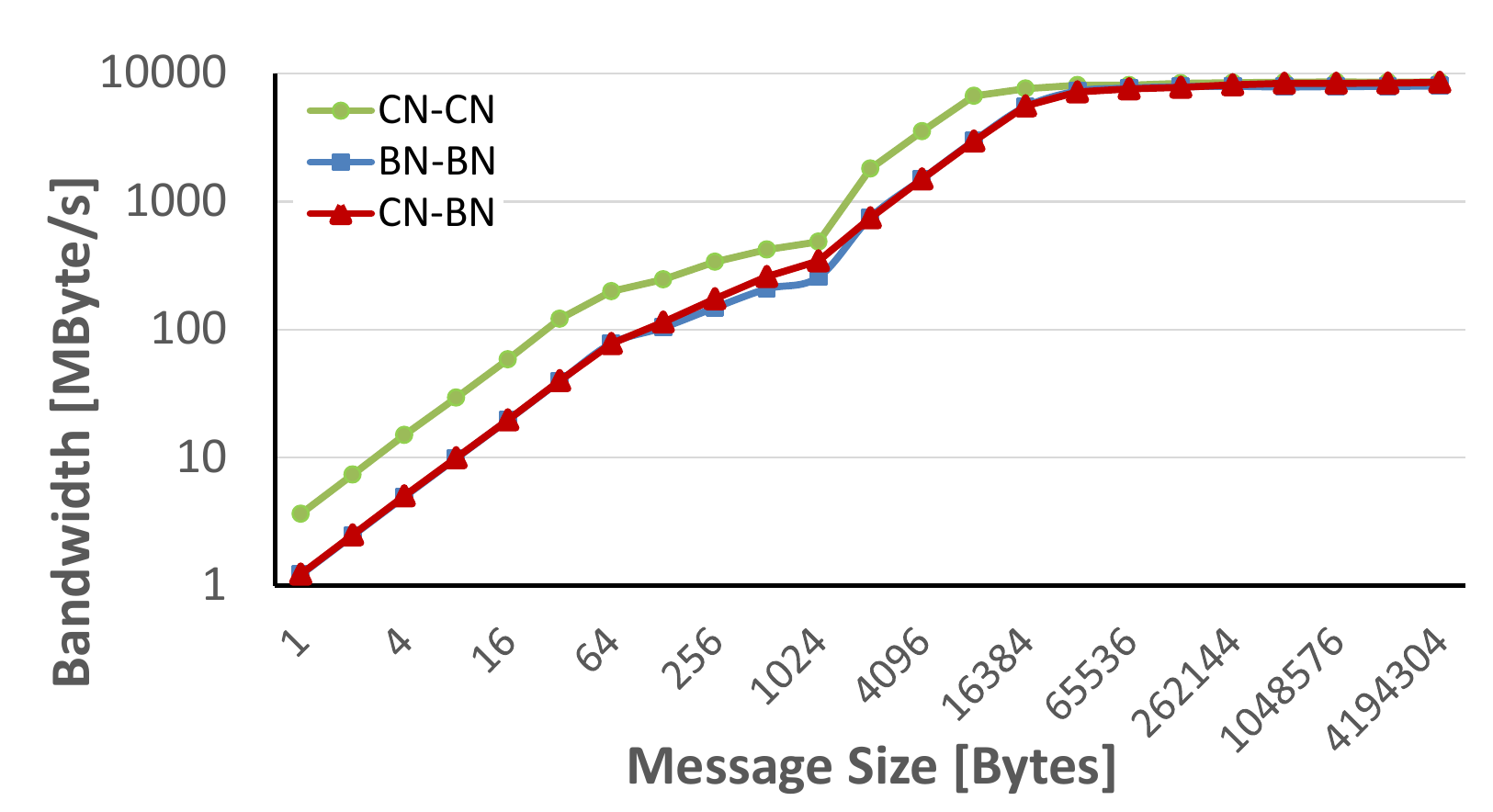}
  %\label{fig:EXTOLLbench_bw}}
  %\hfil
  %\subfloat[Latency]{
  \includegraphics[width=\columnwidth]{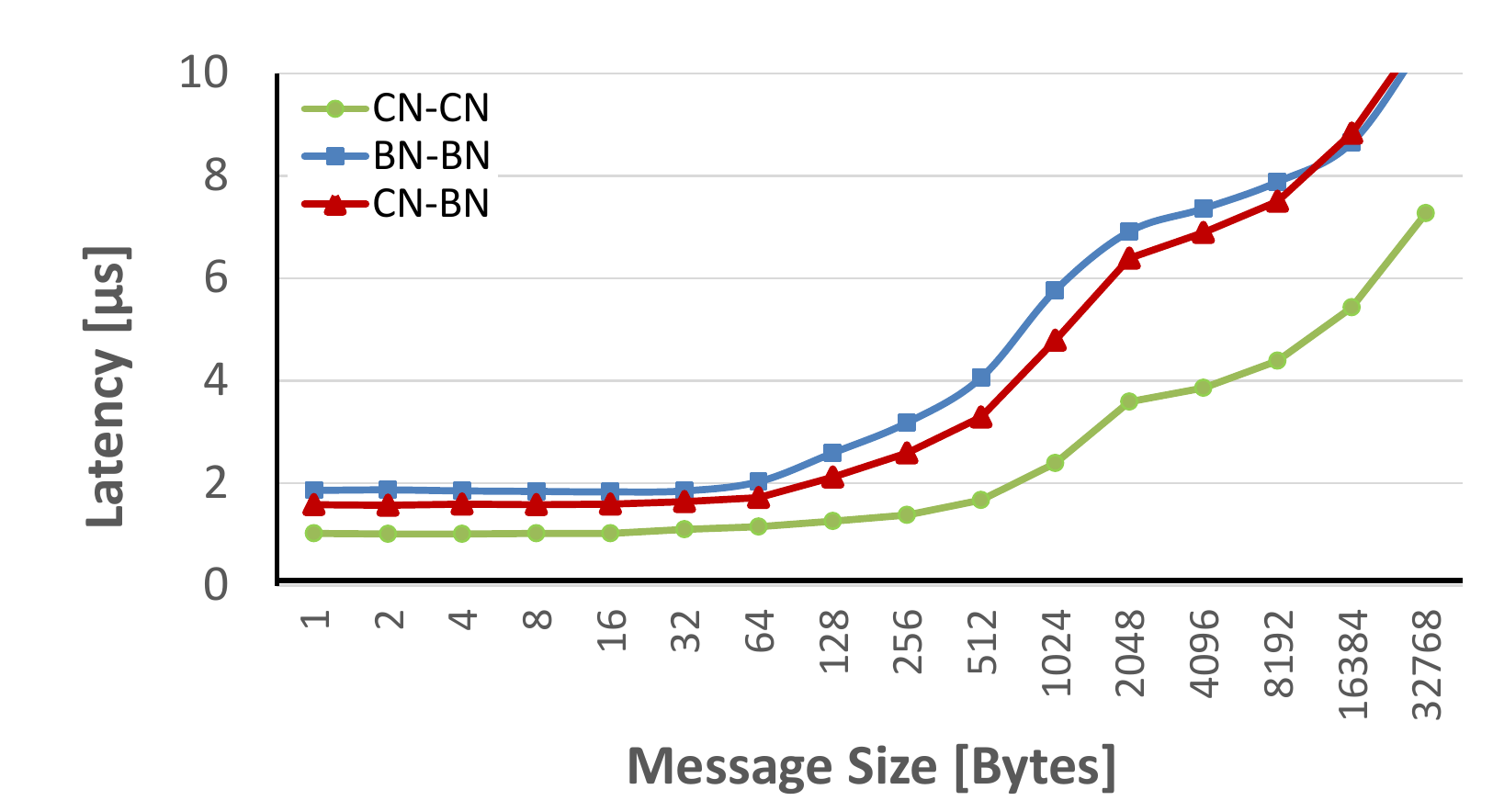}
  %\label{fig:EXTOLLbench_lat}}
  \caption{End-to-end MPI communication (bandwidth and latency)
  measured with ParaStation~MPI on the \mbox{DEEP-ER} prototype.
  (CN: Cluster node; BN: Booster node).
  \textit{CN-CN}: communication between Cluster nodes; 
  \textit{BN-BN}: communication between Booster nodes; 
  \textit{CN-BN}: communication between a Cluster and a Booster node.}
  \label{fig:extoll_bench} 
\end{figure}

A uniform high-speed Tourmalet~A3 EXTOLL fabric  
runs across Cluster and Booster, connecting 
them each internally, between each other, and to the central 
storage. Bandwidth and latency measured by end-to-end 
MPI communications between the different kinds of nodes 
are displayed in figure~\ref{fig:extoll_bench}. For small 
message sizes communication is more efficient between the Cluster 
nodes due to the higher single thread performance of the 
Intel Xeon processors, compared to KNL. 
For large messages communication performance between all 
kinds of nodes is limited by fabric bandwidth. 

%\begin{figure}[!htbp]
%  \centering
%  \includegraphics[width=0.9\columnwidth]{pdf/extoll_bench.pdf}
%  \caption{\label{fig:extoll_bench} End-to-end MPI bandwidth \textit{(a)} 
%  and latency \textit{(b)} measured with ParaStation~MPI on 
%  the \mbox{DEEP-ER} prototype. 
%  \textit{CN-CN}: communication between Cluster nodes; 
%  \textit{BN-BN}: communication between Booster nodes; 
%  \textit{CN-BN}: communication between a Cluster and a Booster node.}
%\end{figure}

\begin{table}[t]\scriptsize
  \centering
    %\begin{tabular}{p{0.7in} p{1.1in} p{1.1in}}
   \caption{\label{tab:hwconfig} Hardware configuration of the \mbox{DEEP-ER} prototype.}
   \begin{tabular}{l l l}
      \hline
      {\bf Feature} & {\bf Cluster} & {\bf Booster}  \\
      \hline \hline
      Processor & Intel Xeon~E5-2680~v3 & Intel Xeon Phi 7210 \\
      Microarchitecture & Haswell & Knights Landing (KNL) \\
      Sockets per node	& 2 & 1 \\
      Cores per node & 24 & 64 \\
      Threads per node & 48 & 256\\
      Frequency  & 2.5~GHz & 1.3~GHz \\
      \hline
      Memory (RAM) & 128~GB & 16~GB -- MCDRAM \\
      & & 96~GB -- DDR4 \\
      NVMe capacity	& 400~GB & 400~GB \\
      % Node architecture & SuperMicro & Intel Adams Pass \\
      \hline
      Interconnect & EXTOLL Tourmalet A3 & EXTOLL Tourmalet A3 \\
      Max. link bandwidth & 100~Gbit/s & 100~Gbit/s \\
      MPI latency\footnotemark 	& 1.0~$\mu$s & 1.8~$\mu$s \\
      % Topology & 3D torus & partial torus \\
      \hline
      Node count & 16 & 8 \\
      Peak performance & 16~TFlop/s & 20~TFlop/s \\
      \hline
    \end{tabular}%
\end{table}%

\footnotetext{Note: The larger MPI latency on the Booster
      is due to the lower single thread performance of the Xeon Phi
      processor. It results from its different micro-architecture in
      combination with the reduced clock frequency compared to
      standard Xeon processors.}

%
%\subsubsection{Memory hierarchy}
%\label{sec:mem}
The \mbox{DEEP-ER} prototype is enhanced by advanced 
memory technologies. A multi-level memory 
hierarchy has been built providing a total memory 
capacity of 8~TBytes. This enables the implementation of 
innovative I/O and resiliency techniques. 
%(see sections~\ref{sec:io} and \ref{sec:resil}). 

Each node of the \mbox{DEEP-ER} prototype (in both Cluster 
and Booster) features a non-volatile memory (NVM) device
for efficient buffering of I/O and checkpointing. 
The chosen technology is Intel's DC~P3700 NVMe device, an replacement
for SSD with 400~GByte capacity that provides high speed, 
non-volatile local memory, directly attached to the node via 4~lanes 
of PCIe~gen3. NVMe is a new standard providing APIs and interfaces for
this direct connection in the server market. On the long run,
it aims at replacing today's standard interfaces like SATA.

%
%\subsubsection{Network Attached Memory (NAM)}
%\label{sec:nam}

\mbox{DEEP-ER} has also introduced an innovative
memory concept: the \textit{network attached memory} 
(NAM)~\cite{schmidtPhD}. 
It combines Hybrid Memory Cube (HMC) 
devices with a state-of-the-art Xilinx Virtex~7 FPGA 
and exploits the remote DMA capabilities of the EXTOLL fabric.
The latter enables access to remote memory resources without 
the intervention of an active component (as a CPU) 
on the remote side.
In this way a
high-speed memory device is created, which is directly attached 
to the EXTOLL fabric and therefore globally accessible by all 
nodes in the system.

The \mbox{DEEP-ER} prototype holds two NAM devices with a capacity 
of 2~GBytes each. This relatively small size is due to current 
HMC technology limitations. Future implementations potentially increase 
capacities and may trigger a rethinking of memory architectures 
for HPC and data analytics. In fact, the NAM concept is being 
further investigated by the successor project \mbox{DEEP-EST}.

\section{Software Environment}
\label{sec:sw}
The guiding principle in the development of software for 
the Cluster-Booster concept has been
to stick, as much as possible, to standards and well 
established APIs.
The specific software features required to operate
Cluster and Booster together as a single system are implemented in
the lower layers of the software stack and are as transparent as
possible to application developers. 
Thus, they experience the same software environment 
as on any other current HPC system and do not have to deal with the 
underlying hardware complexity. Furthermore, their codes stay
portable and keep the capability to run out-of-the-box on this new 
kind of platform as well as on any other HPC system. Thus, application
developers are not required to maintain yet another branch of their
application's source-code but just have to add corresponding pragmas.

\subsection{Programming Environment}
\label{sec:progenv}
The ParaStation~MPI library has been specifically optimized
to efficiently run within both, Cluster and Booster, and across
them. In particular, MPI programs can run solely on the Cluster 
(without employing any Booster node), solely on the Booster (without
using any Cluster node), or run distributed among both kinds of nodes.
For the latter case, enhancements have been done in ParaStation 
implementing a heterogeneous,
\textit{global MPI} by exploiting semantic concepts long existing in
the MPI-standard. In particular, the MPI-2 function 
\texttt{MPI\_Comm\_spawn} realises the offloading mechanism, 
which allows to spawn groups of processes from Cluster to Booster 
(or vice-versa). At the same time the global MPI provides an efficient 
way of exchanging data between the two parts of the system~\cite{eicker:CCPE}.

\begin{figure}
  \centering
  \includegraphics[width=0.9\columnwidth]{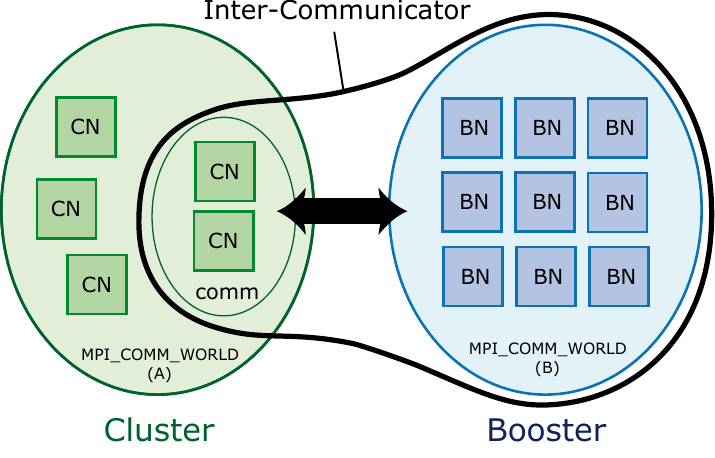}
  \caption{\label{fig:MPI_Comm_spawn}\texttt{MPI\_Comm\_spawn} 
  	schematics, describing an application starting on the 
  	Cluster and offloading a part of its code to the Booster.
  	(CN: Cluster node; BN: Booster node).
	}
\end{figure}

As sketched in figure~\ref{fig:MPI_Comm_spawn},
\texttt{MPI\_Comm\_spawn} is a collective operation performed 
by a (sub-)set of application processes running on 
either Cluster or Booster. The call requires as input the 
name of the binary to run and the number of processes to be started. 
It returns an inter-communicator, providing a connection 
handle to the children. Each child calls then \texttt{MPI\_Init}, 
as usual, and gets access to the other end of the inter-communicator via 
\texttt{MPI\_Get\_parent}. Both parts of the applications 
--~the original main part and the 
offloaded part~-- have their own \texttt{MPI\_COMM\_WORLD}s 
providing full MPI functionality on either side, 
and are connected to each other via inter-communicators.

\subsection{OmpSs abstraction layer}
\label{sec:ompss}
For a programmer, directly employing the 
\texttt{MPI\_Comm\_spawn} functionality requires coordinating 
and managing two or more sets of parallel MPI processes. 
This includes explicitly exchanging the necessary data between 
both sides 
of the system. This approach may become cumbersome 
for large and complex applications. To reduce the porting 
effort, an abstraction layer employing the global~MPI 
has been implemented already in the DEEP project. It enables 
application developers to offload large, complex tasks by 
simply annotating via \texttt{pragmas} these parts of their 
code that shall run on a different part of the system.

This abstraction layer is based on the OmpSs 
data-flow programming model~\cite{duran, ompss:web}: 
an OpenMP~4.0-like environment 
exploiting task-level parallelism and supporting asynchronicity, 
heterogeneity and data movement. With OmpSs, an application's 
code is annotated using OpenMP-like pragmas that indicate data 
dependencies between the different parts of the program. 
Taking these dependencies into account, the OmpSs runtime 
decides on the sequence of tasks and whether concurrent 
execution is allowed, creating a task dependency graph 
at run-time. All this information is used to schedule the 
tasks on the available hardware resources. 

In order to support the DEEP offloading functionality
OmpSs has been extended by an additional pragma to 
indicate the offload of a large compute-task including communication 
from Cluster to Booster, or vice-versa~\cite{sainz}. 
The pragma also holds information on data dependency and by this means
it enables the 
OmpSs source-to-source compiler 
to insert all necessary MPI calls and to pass the resulting sources to
the native tool-chain of the specific part of the system creating
the binaries to be executed.

\subsection{I/O}
\label{sec:io}
The non-volatile memory of the \mbox{DEEP-ER} prototype 
(see~\ref{sec:sysconf}) is used as the foundation of a scalable 
I/O infrastructure. The resulting I/O software platform 
combines the parallel I/O library SIONlib~\cite{Frings:4447} 
with the  
parallel file system BeeGFS~\cite{beegfs}. Together, they enable
the efficient 
and transparent use of the underlying hardware and provide 
the functionality and performance required by data-intensive 
applications and multi-level checkpointing-restart techniques.

The I/O library SIONlib acts as a concentration-layer enabling
applications utilizing task-local I/O to efficiently use the
underlying parallel file system. SIONlib bundles all data locally
generated by applications, and stores it into one or very few
large files, that the file system can easily manage. Furthermore, in
\mbox{DEEP-ER} SIONlib bridges between the I/O and resiliency
components of the software stack. It is used to copy local checkpoints
into the NVM of a companion (\textit{buddy}) node for redundancy, and
to efficiently store checkpoint-data in the global file system. Both
functions work in combination with the scalable checkpointing/restart
library SCR (see section~\ref{sec:resil}).

The file system utilized in \mbox{DEEP-ER} is 
BeeGFS. It provides a solid, common basis 
for high-performance, parallel I/O operations. Advanced 
functionalities, such as a local cache layer in the file system, 
have been added to BeeGFS during the \mbox{DEEP-ER} project.
The cache domain --~based on BeeGFS on demand (BeeOND)~\cite{beeOnd}~-- 
stores data in fast node-local non-volatile memory devices 
and can be used in a synchronous or asynchronous mode. 
This speeds up the applications' I/O operations and reduces 
the frequency of accesses to the global storage, increasing 
the overall scalability of the file system. The corresponding results
will be discussed in detail in an up-coming article~\cite{IOpaper}.

\subsection{Resiliency}
\label{sec:resil}
The \mbox{DEEP-ER} project has adopted an improved 
user-level application checkpoint-restart approach, 
combining it with a task-based resiliency strategy.

%\subsubsection{Checkpoint/restart}
%\label{sec:chkp}
The Open Source Scalable Checkpoint-Restart 
library (SCR) offers a flexible interface 
for applications to perform checkpoints and to restart from 
them in case of failure~\cite{moody:2010}. The user only needs to call SCR 
and indicate the data required by the application 
to restart execution. This library keeps a database 
of checkpoints and their locations in preparation for 
eventual re-initialization. In \mbox{DEEP-ER}, SCR has been extended 
to decide where and how often checkpoints are performed, 
based on a failure model of the \mbox{DEEP-ER} 
prototype. 

The OmpSs programming model has been also extended by
three new resiliency features. Input data of the OmpSs tasks 
can be saved into main memory
before starting them, such that they can be restarted 
in case of failure. Alternatively, the input dependencies 
of a task can be used by OmpSs to fast-forward a re-started 
application to the latest check-point.
Tasks offloaded from Cluster to Booster (or vice-versa)  
can also be restarted without loosing the work that
has been performed in parallel by other OmpSs tasks.

\section{Application results}
\label{sec:results}
Several real-world HPC applications have been
used to steer and evaluate the design of the 
hardware and software developments in the DEEP projects. 
To properly represent the typically 
broad user portfolio of large-scale computer centres,
the chosen co-design applications come 
from a wide range of scientific areas, including astrophysics,
neuroscience, seismic imaging, climate science, computational
fluid dynamics, molecular dynamics, etc. 

The role of these applications in the project is two-fold: 
on the one hand, their requirements have provided co-design 
input to fix the characteristics of hardware and software 
components; on the other hand, the codes have evaluated the
project developments by running different uses cases on the 
\mbox{DEEP-ER} prototype. 

This paper focuses on the distribution of an application 
over both parts of the Cluster-Booster architecture, tested
with the Space Weather application xPic. This code  
has been chosen for two reasons: it displays best the effect of 
partitioning an application between Cluster and Booster, 
and it is the code with which the most exhaustive benchmark-tests 
of this scenario have been performed until now.
Other applications tested on the \mbox{DEEP-ER} 
prototype are of rather monolithic nature, meaning that they run
either on the Cluster or the Booster, alone. Further
heterogeneous simulation-workflows are being adapted to the 
concept in the recently started \mbox{DEEP-EST} project. 
Thorough experimental results will be presented in future publications.

\subsection{Application description and structure}
\begin{figure}[h!]
  \centering
  \includegraphics[width=0.8\columnwidth]{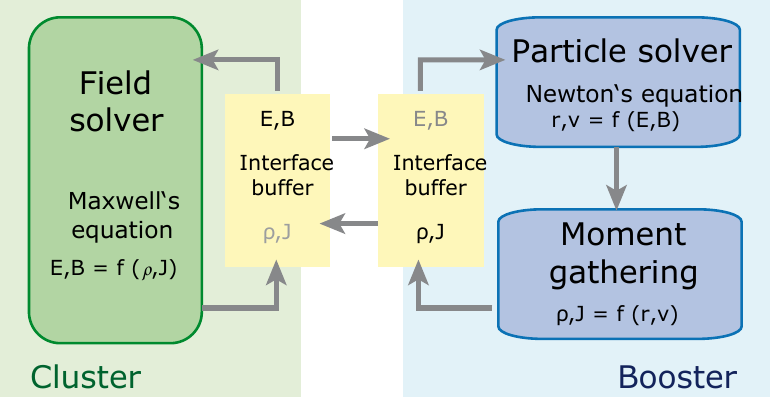}
  \caption{xPic workflow}
  \label{fig:xPic}
\end{figure}

  xPic is a Particle-in-Cell (PIC)
  simulation code from KU~Leuven (Katholieke Universiteit Leuven) to 
  forecast space weather events with the potential
  to harm spacecraft electronics, disturb GPS signals, or 
  even damage the electrical infrastructure on Earth. 
  It simulates the plasma produced in solar eruptions using the 
  Implicit Moment Method~\cite{ipic3d}. 
  Like most PIC 
  codes, xPic consists of two parts, a particle solver 
  and a field solver. The particle solver calculates the 
  motion of charged particles in response to the 
  electromagnetic field and collects statistical information about their charge density, velocity distribution and the corresponding electric current (called moment gathering); 
  the field solver computes the electromagnetic field 
  evolution in response to the particle movement. The workflow 
  of xPic is presented in figure~\ref{fig:xPic}. Here, 
  the color-coding employed along the paper has been
  kept: code-parts best suited for the Cluster are 
  marked in \textit{green}, while those in \textit{blue} 
  fit best on the Booster side of the \mbox{DEEP-ER} prototype.
  
\subsection{Distributing an application between Cluster and Booster}
The applications tested in the \mbox{DEEP-ER} prototype 
employed two alternative approaches to take advantage of 
the Cluster-Booster system: (1) launching MPI processes in 
one of the modules and spawning children MPI processes on the 
second module (see section~\ref{sec:progenv}); and (2) 
using OmpSs pragmas to offload computing zones from the 
Cluster to the Booster (section~\ref{sec:ompss}). 
The developers of the xPic code explored both approaches at the 
beginning of the DEEP project but finally decided to go for the first 
of them, due to their personal larger experience in MPI programming.
The xPic code has then been divided in a particle 
solver that runs on Booster nodes and a field solver 
that runs on Cluster nodes. The application can also run 
in traditional architectures, by executing particle and field solver
on the same kind of nodes.

The listing~1 in figure~\ref{fig:listings} shows the main loop of xPic
in its original configuration. 
The field and particle solvers are labelled \texttt{fld} 
and \texttt{pcl} respectively. The solver calculates the 
electric (E) and magnetic (B) fields, while the particle 
solver performs the particle movement and the moment gathering. 
The functions \texttt{cpyFromArr} and \texttt{cpyToArr} 
move information between the solvers and the 
interface buffer shown in figure~\ref{fig:xPic}. 

%\lstinputlisting[language=C++,caption={main loop}]{mainloop.txt}

In the Cluster-Booster mode, the main loop is divided into 
two files, one containing the Cluster routines, and the 
second containing the Booster routines. In practice, the 
developer creates two copies of the main file and erases the 
\texttt{fld} calls in the Booster copy, and the \texttt{pcl} 
calls in the Cluster copy. Finally, after each \texttt{cpy} 
call, data is moved between the two solvers.

Figure~\ref{fig:listings} show the original main loop, together
with the newly defined main loops for Cluster (listing~2) and Booster (listing~3). 
Differences are highlighted in green for the Cluster parts,
and in blue for the Booster parts.

\begin{figure*}
\begin{flushleft}
\begin{minipage}{.32\textwidth}
%\minipage{.32\textwidth}
\begin{lstlisting}[language=C++,
	caption={Original \texttt{main} loop}]

for (auto i=beg+1; i <= end; i++){
  fld.solver->calculateE(); 
  fld.cpyToArr_F();



  pcl.cpyFromArr_F();
  for (auto is=0; is<nspec; is++) {
    pcl.species[is].ParticlesMove();
    pcl.species[is].ParticleMoments();
  }
  pcl.cpyToArr_M();



  fld.solver->calculateB(); 
  fld.cpyFromArr_M();
}
(*@ @*)
\end{lstlisting}
	\label{list:OldMain}
\end{minipage}
%\endminipage
\hfill
\begin{minipage}{.29\textwidth}
\begin{lstlisting}[language=C++, numbers=right, numbersep=5pt,
	caption={Cluster \texttt{main} loop}]
(*@\textcolor{ForestGreen}{\textbf{\#ifdef} \_\_CLUSTER\_\_} @*)
for (auto i=beg+1; i <= end; i++){
  fld.solver->calculateE(); 
  fld.cpyToArr_F();
  (*@\textcolor{ForestGreen}{ClusterToBooster();} @*)
  (*@\textcolor{ForestGreen}{// Auxiliary computations} @*)
  (*@\textcolor{ForestGreen}{ClusterWait();} @*)






  (*@\textcolor{ForestGreen}{BoosterToCluster();} @*)

  (*@\textcolor{ForestGreen}{BoosterWait();} @*)
  fld.solver->calculateB(); 
  fld.cpyFromArr_M();
}
(*@\textcolor{ForestGreen}{\textbf{\#endif}} @*)
\end{lstlisting}
	\label{list:ClusterMain}
\end{minipage}
\hfill
\begin{minipage}{.32\textwidth}
\begin{lstlisting}[language=C++,
	caption={Booster \texttt{main} loop}]
(*@\textcolor{RoyalBlue}{\textbf{\#ifdef} \_\_BOOSTER\_\_} @*)
for (auto i=beg+1; i <= end; i++){


  (*@\textcolor{RoyalBlue}{ClusterToBooster();} @*)

  (*@\textcolor{RoyalBlue}{ClusterWait();} @*)
  pcl.cpyFromArr_F();
  for (auto is=0; is<nspec; is++) {
    pcl.species[is].ParticlesMove();
    pcl.species[is].ParticleMoments();
  }
  pcl.cpyToArr_M();
  (*@\textcolor{RoyalBlue}{BoosterToCluster();} @*)
  (*@\textcolor{RoyalBlue}{// I/O and auxiliary computations} @*)
  (*@\textcolor{RoyalBlue}{BoosterWait();} @*)


}
(*@\textcolor{RoyalBlue}{\textbf{\#endif}} @*)
\end{lstlisting}
	\label{list:BoosterMain}
\end{minipage}
\end{flushleft}
\caption{Listings showing the \texttt{main} loop in 
	the original (Listing 1) and new  
	xPic application. In the new version 
	the loop is distributed between Cluster (Listing~2) 
	and Booster (Listing~3). Cluster-to-Booster MPI 
	communications have been added in blue and green.  
	Lines 6 and 15 represent computations that can be done 
	while the non-blocking communications are performed.}
	\label{fig:listings}
\end{figure*}

The functions \texttt{ClusterToBooster} and 
\texttt{BoosterToCluster} perform the MPI communications 
between the two modules. These are non blocking, and 
allow to overlap with non critical operations, like the 
computations of particle and field energy, the post-processing 
of data, and writing output files. Communications are performed 
using the \texttt{INTERCOMM} communicator created at the 
initialisation of the code with the \texttt{MPI\_Comm\_spawn} 
routine. Listing~4 shows how the function 
\texttt{BoosterToCluster()} uses these MPI communications. 

%\begin{minipage}[c]{0.95\columnwidth}
\begin{lstlisting}[language=C++,caption={Booster to Cluster MPI communication}]
(*@ \textcolor{RoyalBlue}{\textbf{\#ifdef} \_\_BOOSTER\_\_} @*)
  (*@ \textcolor{RoyalBlue}{MPI\_Issend(Rho,..., INTERCOMM, nextSreq());} @*)
(*@ \textcolor{RoyalBlue}{\textbf{\#endif }} @*)
(*@ \textcolor{ForestGreen}{\textbf{\#ifdef} \_\_CLUSTER\_\_} @*)
  (*@ \textcolor{ForestGreen}{MPI\_Irecv (Rho,..., INTERCOMM, nextRreq());} @*)
(*@ \textcolor{ForestGreen}{\textbf{\#endif}} @*)
\end{lstlisting}
%\end{minipage}

The compilation script generates two executables, one containing 
the \texttt{\_\_BOOSTER\_\_} code and the second containing 
the \texttt{\_\_CLUSTER\_\_} code. At launch time, the execution 
script calls the Booster code, and this in turn performs a 
spawn with the name of the Cluster executable. ParaStation and the 
scheduler detect this call and distribute the child binaries 
in the correct locations in the Cluster.
%\textcolor{green}{JA: is this last sentence accurate?}

\subsection{Benchmarking results}
\begin{figure}[h!]
  \centering
  \includegraphics[width=0.9\columnwidth]{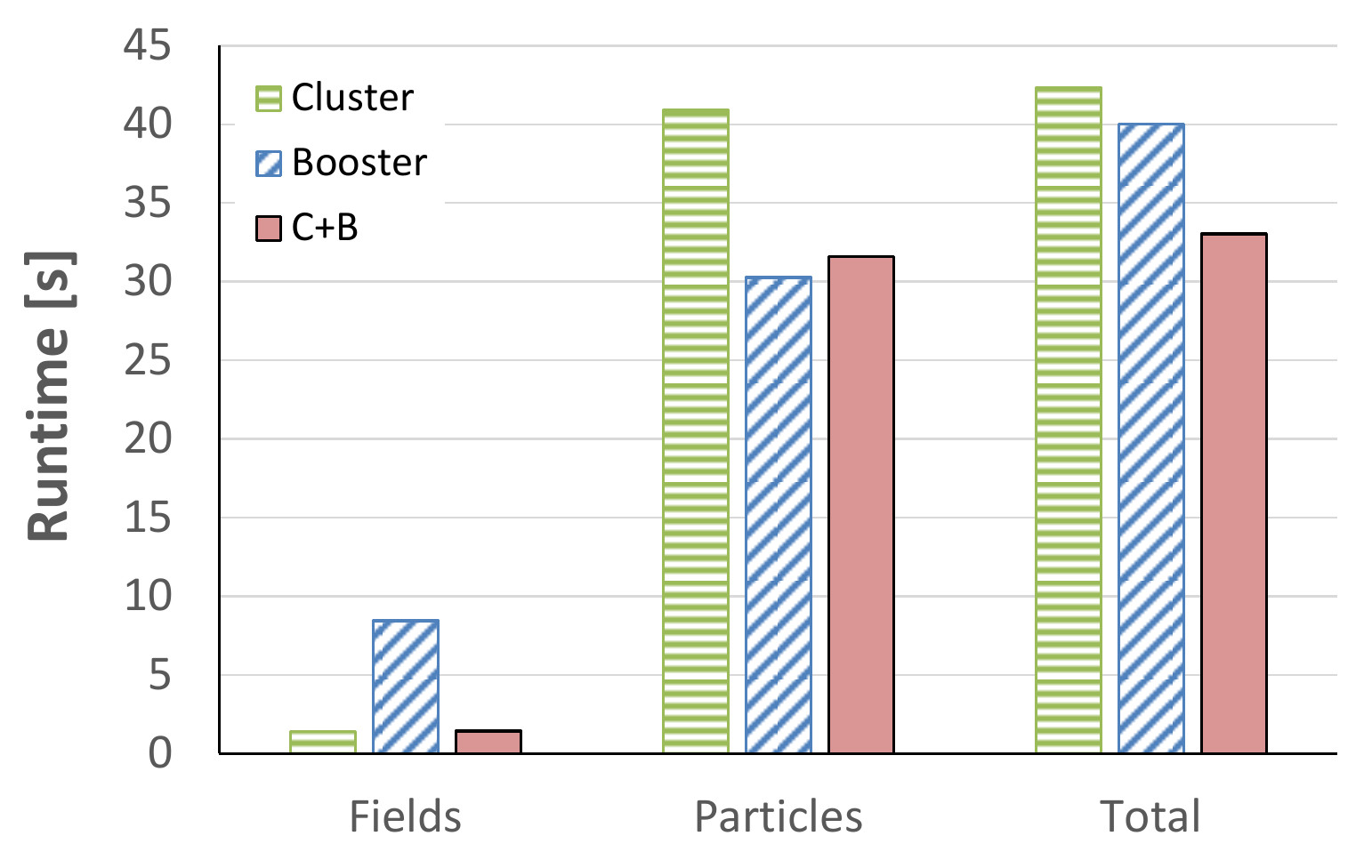}
  \caption{Runtime of xPic and its constituents: running
  	both solvers on the Cluster, both on the Booster,
  	and using the Cluster-Booster mode (labelled \textit{C+B}).
  	In the latter case the field solver runs on the Cluster and 
  	the particle solver on the Booster.}
  \label{fig:xPic_CBdiv}
\end{figure}

Figure \ref{fig:xPic_CBdiv} illustrates how the application xPic 
profits from distributing its two solvers (fields and particles)  
over the Cluster-Booster architecture described 
in section \ref{sec:CBconcept}. The results in this figure have 
been obtained using single Cluster and Booster nodes. Each 
solver uses a hybrid MPI+OpenMP code. The experimental 
setup is summarized in table~\ref{tab:xPic_CBMsetup}.
In this case, running only on the Cluster means 
executing the particle 
solver on one Cluster node first and, once finished, 
using the same node for the field solver. The total execution
time is the sum of the time employed by both solvers. 
The same applies to the case that 
uses only one Booster node. The Cluster-Booster mode 
(labelled \textit{C+B}) runs the field solver 
on one Cluster node and the particle solver on one 
Booster node. The total execution time is here the sum of both
parts and includes the overhead  
due to the MPI communication between them.

\begin{table}[!htbp]\scriptsize
  \centering
    %\begin{tabular}{|p{5cm}|p{5cm}|}
   \caption{\label{tab:xPic_CBMsetup} xPic experiment setup in the 
  Cluster-Booster architecture evaluation measurements.}   		 
  \begin{tabular}{|l|l|}
      \hline
      Number of cells per node & 4096\\
      \hline
      Number of particles per cell & 2048\\
      \hline
      Compilation flags & -openmp, -mavx (Cluster), \\
      & -xMIC-AVX512 (Booster) \\
      \hline 
    \end{tabular}
\end{table}

The field solver matches best to the Cluster side, 
since this code-part is not highly parallel and requires 
substantial and frequent global communication. Accordingly, 
running the field solver on the Cluster (Haswell processors) 
is 6$\times$ faster than on the Booster (KNL processors). 
The highly parallel particle solver, on the other 
hand, moves billions of particles independently with 
almost no long-range communication.  
It turns out to be naturally suited to the Booster, where it 
runs about 1.35$\times$ faster than on the Cluster.
Point-to-point communication is done between the field solver 
and particle solver (i.e. between Cluster and Booster) 
and constitutes only a small fraction 
(3\% to 4\% overhead per solver)
of the total application communication. 

Thus, the Cluster-Booster architecture allows matching the
intrinsic structure of xPic to the hardware, i.e.~running the field
solver on the Cluster and the particle
solver on the Booster. This distributed mode results in 
a 1.28$\times$ performance gain of the overall application,  
when compared
to running the full code using only the Cluster. Comparing 
to an execution on the Booster alone, still a 1.21$\times$ 
performance gain of the Cluster-Booster (C+B) mode is achieved.

\begin{figure}[!tbp]
  \centering
  %\subfloat[Runtime]{
  \includegraphics[width=0.9\columnwidth]{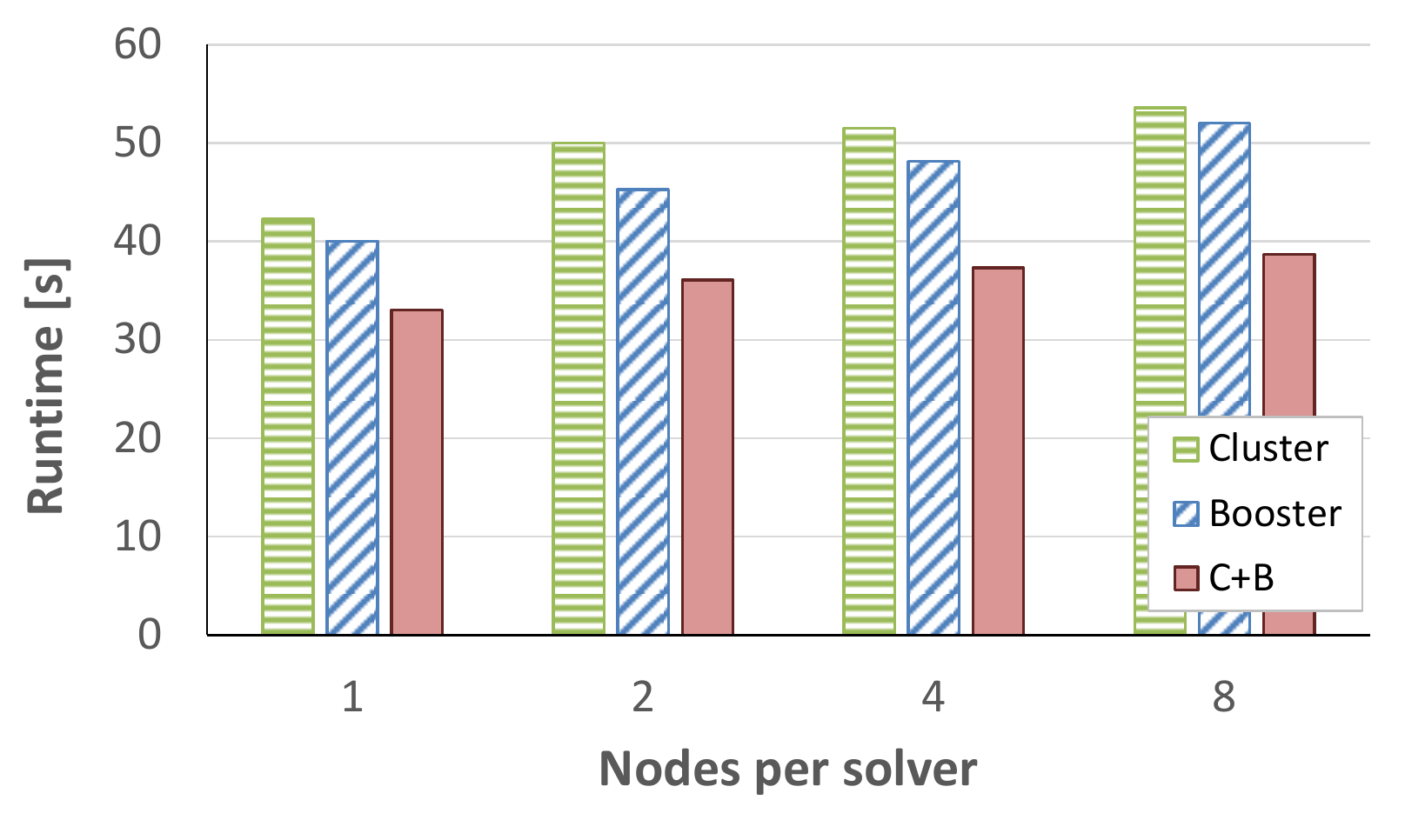}
  %\label{fig:xPic_CBdiv_scale_runtime}}
  %\hfil
  %\subfloat[Parallel efficiency]{
  \includegraphics[width=0.9\columnwidth]{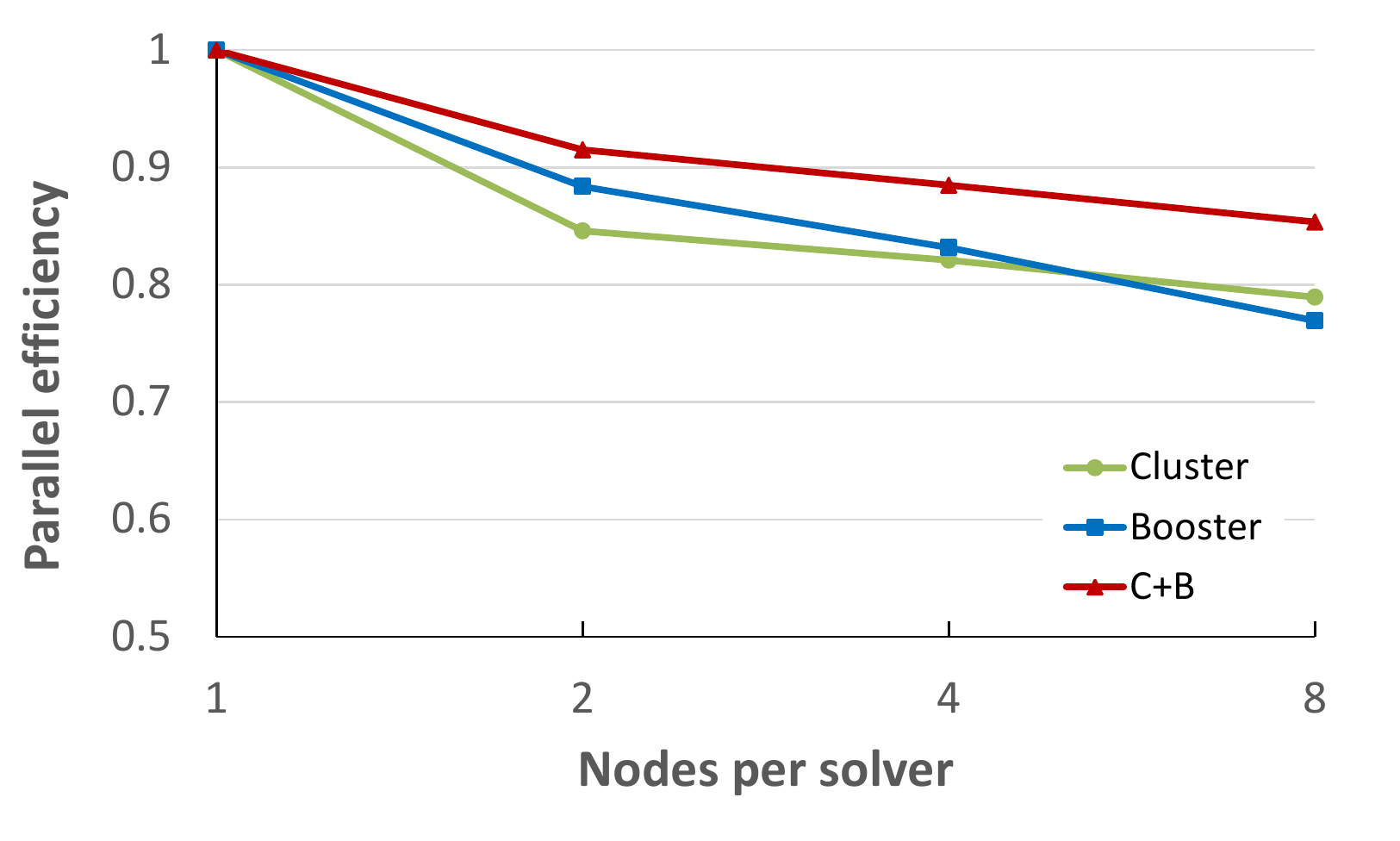}
  %\label{fig:xPic_CBdiv_scale_effic}}
  \caption{\label{fig:xPic_CBscale} Scaling results (runtime 
  and efficiency) with xPic: 
   	only Cluster, only Booster, 
   	and using the Cluster-Booster mode (\textit{C+B}).}
\end{figure}

Scaling results for the three scenarios (only Cluster, only Booster,
and Cluster-Booster mode) are presented in figure \ref{fig:xPic_CBscale}. 
The plots indicate that the performance gain of the C+B mode 
increases with the number of nodes. In the largest experiment
possible on the \mbox{DEEP-ER} prototype (8 nodes), the distributed
code runs 1.38$\times$ faster than using only the Cluster, 
and 1.34$\times$ faster
than on the Booster alone. The C+B mode also achieves a better 
parallel efficiency (85\%) than using the Cluster (79\%) and 
Booster (77\%) as stand-alone systems.

\section{Related work}
\label{sec:relwork}
This paper presents a different approach for the integration of heterogeneous
resources within a HPC system. In fact, the actual idea is similar to
the concept behind the development of the Quadrics Supercomputing
World's PQE~2000 system in the late 1990s\cite{PQE2000}. Here the role
of the Cluster as a more general purpose system was filled by a Meiko
CS2 system utilizing SPARC processors and a proprietary interconnect
build by Meiko. The part now named Booster was planned to be realized
by a highly scalable APE Mille MPP system that was based on the APE
series of machines originally designed for lattice QCD calculations.
Nevertheless, at that time the idea did not make it to the market.

The original Cluster-Booster prototype of the DEEP project was
challenging to realize since the first generation of Intel Xeon Phi
processors were not designed to run as stand-alone processors.  With
the advent of Xeon Phi processors of the KNL generation also major
hardware vendors like Cray offer systems that integrate Intel Xeon and
Intel Xeon Phi processors into a single system. Examples for larger
systems of this type are the Cori system at NERSC~\cite{Cori} based on
Cray's XC40 series or the Trinity system at LANL~\cite{Trinity} based
on Cray XC30. However, until now there is no indication that these
systems will be used in the same fashion that is presented in this
paper, i.e.~by running applications across both type of processor
architectures at the same time, utilizing \texttt{MPI\_Comm\_spawn}
or similar calls in the MPI standard. In fact,
\texttt{MPI\_Comm\_spawn} was not supported by Cray's MPI until
recently.

In a more general sense the integration of heterogeneous resources
into a single system is available in many large-scale HPC system. 
They might have large memory nodes in order to support
applications with the need for larger amounts of memory, although
those applications are usually restricted to a single class of nodes
alone. More in the sense of the approach presented in this paper are
visualization nodes within a large scale supercomputer used for online
visualization. Here both classes of nodes are used at the same time,
one for running the actual application, the other in order to tap data
from the running simulation and to derive graphical representation of
these data. Nevertheless, in both cases the heterogeneity of the nodes
is restricted to a different amount of memory or additional hardware
like GPUs but typically leaving the processor architecture
untouched. Furthermore there is no spawning of additional processes
via \texttt{MPI\_Comm\_spawn}. Instead, communication between the different
application parts (simulation and visualization) is done by different
measures.

For the concept of NAM a similar approach is realized by the RAM Area
Network developed by Kove in its xpd appliance~\cite{Kove}. While NAM
directly attaches HMC memory to the EXTOLL interconnect, Kove utilizes
standard DRAM DIMMs and multiple InfiniBand HCA in order to realize
larger capacity and higher bandwidth. The main difference between the
two concepts is that the xpd appliance still requires a standard
processor while for the NAM all functionality is integrated in a
single FPGA.

\section{Conclusions and Outlook}
\label{sec:concl}
The DEEP projects have introduced several
hardware and software innovations to improve the capabilities 
of today's HPC systems, addressing several of 
the Exascale challenges. In particular, an
innovative architecture concept has been introduced,
which provides the applications with full flexibility on how 
to exploit different kinds of computing resources. 

The Cluster-Booster architecture integrates
heterogeneous resources at the system level, instead of the
node level. The Booster (a cluster
of many-core processors or accelerators) is attached to
a Cluster (a system of general purpose processors)
via a high-speed network. 
Application developers have full freedom to decide how they 
distribute their codes over the system and can match the
requirements of their different code parts to the available hardware.

The performance improvement that this approach can provide to 
real-world applications has been demonstrated by the Space 
Weather code xPic. It was able to achieve its results
in shorter time, and with a better parallel efficiency when
distributing the code over Cluster and Booster, than when
running separately on any of them. 

It is important to mention that these results have been
achieved without compromising the portability of the code, 
which regularly runs on other ``standard'' HPC systems. 
This has been achieved by using standard interfaces 
and software components.

The Cluster-Booster architecture, which was first prototyped in
the DEEP projects, has gone into production in the meantime. 
The JURECA Cluster, running at the J\"{u}lich
Supercomputing Center (JSC) in Germany since 2015~\cite{jureca}, 
has been recently accompanied by a KNL-based, 5~PFlop/s Booster,
which is planned to become available to users in Q1/2018. 

The DEEP and \mbox{DEEP-ER} projects 
have been completed and successfully
evaluated by external reviewers. Building on their results, 
the successor project (\mbox{DEEP-EST}) currently 
generalises the Cluster-Booster concept to create a
\textit{Modular Supercomputing architecture}~\cite{Suarez:844072}. It combines any number 
of compute modules (Cluster and Booster are two such modules)
into a unified computing platform. Each compute module is a 
cluster of a potentially large size, tailored to 
the specific needs of a class of applications. 
A high-speed interconnect between the modules and a 
uniform software stack across them enables codes and 
work-flows to run distributed over the
whole machine, matching their specific needs to the available 
computing resources. One of the most
important contributions expected from \mbox{DEEP-EST} is the further
enhancement of resource management software and scheduling strategies
to deal with any number of compute modules. To demonstrate its
capabilities, a hardware prototype consisting of three modules 
will be built. It shall cover the needs of both HPC and 
high performance data analytics (HPDA) workloads.

In parallel to \mbox{DEEP-EST}, JSC is already starting the
implementation of the Modular Supercomputing architecture in a
large-scale production system. The first module of the new Modular
Supercomputing infrastructure will be a general purpose cluster, 
to be deployed in Q2/2018. Its Booster component is planned
for 2019/2020. Further modules will be added in the future, always
aiming at optimally addressing the needs of the wide spectrum of user
communities and applications running at the HPC centre.

\section*{Acknowledgements}
\label{sec:ack}
The authors would like to thank all the people and 
partners involved in the DEEP consortia for 
their engagement and strong commitment
towards the DEEP projects, which led to several of the 
results described in this paper. 
Special gratitude goes to M.\ N\"{u}ssle from EXTOLL~GmbH for 
the MPI benchmarks on the Tourmalet network (figure~\ref{fig:extoll_bench}).

Part of the research presented here has received funding 
from the European Community's Seventh Framework Programme 
(FP7/2007-2013) under Grant Agreement n$^\circ$~287530 (DEEP) 
and 610476 (\mbox{DEEP-ER}), and from the Horizon~2020 
Programme (H2020-FETHPC) under Grant Agreement n$^\circ$~754304 
(\mbox{DEEP-EST}). The present publication reflects only the 
authors' views. The European Commission is not liable for 
any use that might be made of the information contained therein. 

%\newpage

% trigger a \newpage just before the given reference
% number - used to balance the columns on the last page
% adjust value as needed - may need to be readjusted if
% the document is modified later
%\IEEEtriggeratref{8}
% The "triggered" command can be changed if desired:
%\IEEEtriggercmd{\enlargethispage{-5in}}

% references section

% can use a bibliography generated by BibTeX as a .bbl file
% BibTeX documentation can be easily obtained at:
% http://www.ctan.org/tex-archive/biblio/bibtex/contrib/doc/
% The IEEEtran BibTeX style support page is at:
% http://www.michaelshell.org/tex/ieeetran/bibtex/
%\bibliographystyle{IEEEtran}
% argument is your BibTeX string definitions and bibliography database(s)
%\bibliography{IEEEabrv,../bib/paper}
%
% <OR> manually copy in the resultant .bbl file
% set second argument of \begin to the number of references
% (used to reserve space for the reference number labels box)

%\newpage

\bibliographystyle{IEEEtran}
\bibliography{biblio}

% Generated by IEEEtran.bst, version: 1.13 (2008/09/30)
\begin{thebibliography}{10}
\providecommand{\url}[1]{#1}
\csname url@samestyle\endcsname
\providecommand{\newblock}{\relax}
\providecommand{\bibinfo}[2]{#2}
\providecommand{\BIBentrySTDinterwordspacing}{\spaceskip=0pt\relax}
\providecommand{\BIBentryALTinterwordstretchfactor}{4}
\providecommand{\BIBentryALTinterwordspacing}{\spaceskip=\fontdimen2\font plus
\BIBentryALTinterwordstretchfactor\fontdimen3\font minus
  \fontdimen4\font\relax}
\providecommand{\BIBforeignlanguage}[2]{{%
\expandafter\ifx\csname l@#1\endcsname\relax
\typeout{** WARNING: IEEEtran.bst: No hyphenation pattern has been}%
\typeout{** loaded for the language `#1'. Using the pattern for}%
\typeout{** the default language instead.}%
\else
\language=\csname l@#1\endcsname
\fi
#2}}
\providecommand{\BIBdecl}{\relax}
\BIBdecl

\bibitem{DEEPweb}
\BIBentryALTinterwordspacing
(2018) The {DEEP}~projects website. [Online]. Available:
  \url{http://www.deep-projects.eu/}
\BIBentrySTDinterwordspacing

\bibitem{eicker:PARS}
\BIBentryALTinterwordspacing
N.~Eicker and T.~Lippert, ``{A}n accelerated {C}luster-{A}rchitecture for the
  {E}xascale,'' in \emph{PARS '11, PARS-Mitteilungen, Mitteilungen -
  Gesellschaft f\"{u}r Informatik e.V., Parallel-Algorithmen und
  Rechnerstrukturen, ISSN 0177-0454, Nr. 28, Oktober 2011 (Workshop 2011), 110
  - 119}, 2011, record converted from VDB: 12.11.2012. [Online]. Available:
  \url{http://juser.fz-juelich.de/record/19212}
\BIBentrySTDinterwordspacing

\bibitem{eicker:CCPE}
\BIBentryALTinterwordspacing
N.~Eicker, T.~Lippert, T.~Moschny, E.~Suarez, and for~the DEEP~project, ``The
  deep project: An alternative approach to heterogeneous cluster-computing in
  the many-core era,'' \emph{Concurrency and Computation: Practice and
  Experience}, vol.~28, no.~8, pp. 2394--2411, 2016, cpe.3562. [Online].
  Available: \url{http://dx.doi.org/10.1002/cpe.3562}
\BIBentrySTDinterwordspacing

\bibitem{Suarez:844072}
\BIBentryALTinterwordspacing
E.~Suarez, N.~Eicker, and T.~Lippert, ``{S}upercomputer {E}volution at {JSC},''
  ser. Publication Series of the John von Neumann Institute for Computing (NIC)
  NIC Series, vol.~49, NIC Symposium 2018, Jülich (Germany), 22 Feb 2018 - 23
  Feb 2018.\hskip 1em plus 0.5em minus 0.4em\relax Jülich: John von Neumann
  Institute for Computing, Feb 2018, pp. 1 -- 12. [Online]. Available:
  \url{http://juser.fz-juelich.de/record/844072}
\BIBentrySTDinterwordspacing

\bibitem{prabhakaran}
\BIBentryALTinterwordspacing
S.~Prabhakaran, M.~Neumann, S.~Rinke, F.~Wolf, A.~Gupta, and L.~V. Kale, ``A
  batch system with efficient adaptive scheduling for malleable and evolving
  applications,'' in \emph{Proceedings of the 2015 IEEE International Parallel
  and Distributed Processing Symposium}, ser. IPDPS '15.\hskip 1em plus 0.5em
  minus 0.4em\relax Washington, DC, USA: IEEE Computer Society, 2015, pp.
  429--438. [Online]. Available: \url{http://dx.doi.org/10.1109/IPDPS.2015.34}
\BIBentrySTDinterwordspacing

\bibitem{schmidtPhD}
\BIBentryALTinterwordspacing
J.~Schmidt, ``Accelerating checkpoint/restart application performance in
  large-scale systems with network attached memory,'' Ph.D. dissertation,
  Ruprecht-Karls University Heidelberg, Faculty for Mathematics and Computer
  Science, 2017. [Online]. Available:
  \url{http://archiv.ub.uni-heidelberg.de/volltextserver/23800/1/dissertation_juri_schmidt_publish.pdf}
\BIBentrySTDinterwordspacing

\bibitem{duran}
A.~Duran, E.~Ayguad\'{e}, R.~M. Badia, J.~Labarta, L.~Martinell, X.~Martorell,
  and J.~Planas, ``Ompss: A proposal for programming heterogeneous multi-core
  architectures,'' \emph{Parallel Processing Letters}, vol.~21, no.~02, pp.
  173--193, 2011.

\bibitem{ompss:web}
\BIBentryALTinterwordspacing
(2018) Ompss website. [Online]. Available: \url{https://pm.bsc.es/ompss}
\BIBentrySTDinterwordspacing

\bibitem{sainz}
F.~Sainz, J.~Bell\'{o}n, V.~Beltran, and J.~Labarta, ``Collective offload for
  heterogeneous clusters,'' in \emph{2015 IEEE 22nd International Conference on
  High Performance Computing (HiPC)}, Dec 2015, pp. 376--385.

\bibitem{Frings:4447}
\BIBentryALTinterwordspacing
W.~Frings, F.~Wolf, and V.~Petkov, ``{S}calable {M}assively {P}arallel {I}/{O}
  to {T}ask-{L}ocal {F}iles,'' in \emph{Proceedings of the Conference on High
  Performance Computing Networking, Storage and Analysis, Portland, Oregon,
  November 14 - 20, 2009, SC'09, SESSION: Technical papers, Article No. 17, New
  York, ACM, 2009.ISBN 978-1-60558-744-8. - S. 1 - 11}, 2009, record converted
  from VDB: 12.11.2012. [Online]. Available:
  \url{http://juser.fz-juelich.de/record/4447}
\BIBentrySTDinterwordspacing

\bibitem{beegfs}
\BIBentryALTinterwordspacing
(2018) {BeeGFS} website. [Online]. Available:
  \url{https://www.beegfs.io/content/}
\BIBentrySTDinterwordspacing

\bibitem{beeOnd}
\BIBentryALTinterwordspacing
(2018) {BeeGFS On Demand} website. [Online]. Available:
  \url{https://www.beegfs.io/wiki/BeeOND}
\BIBentrySTDinterwordspacing

\bibitem{IOpaper}
\BIBentryALTinterwordspacing
A.~Kreuzer, J.~Amaya, N.~Eicker, R.~Léger, and E.~Suarez, ``{T}he {DEEP}-{ER}
  project: {I}/{O} and resiliency extensions for the {C}luster-{B}ooster
  architecture,'' 2018 IEEE 20th International Conference on High Performance
  Computing and Communications, Exeter (United Kingdom), 28 Jun 2018 - 30 Jun
  2018.\hskip 1em plus 0.5em minus 0.4em\relax IEEE, Jun 2018, pp. 109 -- 116.
  [Online]. Available: \url{http://juser.fz-juelich.de/record/860444}
\BIBentrySTDinterwordspacing

\bibitem{moody:2010}
\BIBentryALTinterwordspacing
A.~Moody, G.~Bronevetsky, K.~Mohror, and B.~R.~d. Supinski, ``Design, modeling,
  and evaluation of a scalable multi-level checkpointing system,'' in
  \emph{Proceedings of the 2010 ACM/IEEE International Conference for High
  Performance Computing, Networking, Storage and Analysis}, ser. SC '10.\hskip
  1em plus 0.5em minus 0.4em\relax Washington, DC, USA: IEEE Computer Society,
  2010, pp. 1--11. [Online]. Available:
  \url{https://doi.org/10.1109/SC.2010.18}
\BIBentrySTDinterwordspacing

\bibitem{ipic3d}
\BIBentryALTinterwordspacing
S.~Markidis, G.~Lapenta, and Rizwan-uddin, ``Multi-scale simulations of plasma
  with ipic3d,'' \emph{Mathematics and Computers in Simulation}, vol.~80,
  no.~7, pp. 1509 -- 1519, 2010, multiscale modeling of moving interfaces in
  materials. [Online]. Available:
  \url{http://www.sciencedirect.com/science/article/pii/S0378475409002444}
\BIBentrySTDinterwordspacing

\bibitem{PQE2000}
P.~Palazzari, L.~Arrcipiani, M.~Celino, R.~Guadagni, A.~Marongiu, A.~Mathis,
  P.~Novelli, and V.~Rosato, ``Heterogeneity as key feature of high performance
  computing: the pqe1 prototype,'' in \emph{Proceedings 9th Heterogeneous
  Computing Workshop (HCW 2000) (Cat. No.PR00556)}, 2000, pp. 17--30.

\bibitem{Cori}
\BIBentryALTinterwordspacing
(2018) Cori website. [Online]. Available:
  \url{http://www.nersc.gov/users/computational-systems/cori/}
\BIBentrySTDinterwordspacing

\bibitem{Trinity}
\BIBentryALTinterwordspacing
(2018) Trinity website. [Online]. Available:
  \url{http://www.lanl.gov/projects/trinity/}
\BIBentrySTDinterwordspacing

\bibitem{Kove}
\BIBentryALTinterwordspacing
(2018) Kove website. [Online]. Available: \url{http://kove.net/xpd}
\BIBentrySTDinterwordspacing

\bibitem{jureca}
\BIBentryALTinterwordspacing
D.~Krause and P.~Th\"{o}rnig, ``{JURECA}: {G}eneral-purpose supercomputer at
  {J}\"{u}lich {S}upercomputing {C}entre,'' \emph{Journal of large-scale
  research facilities}, vol.~2, p. A62, 2016. [Online]. Available:
  \url{http://juser.fz-juelich.de/record/807073}
\BIBentrySTDinterwordspacing

\end{thebibliography}

\end{document}